\documentclass{DISproc}
\usepackage{bm}
\begin{document}
\newcommand{\tr}{{\rm tr}} 
\title{Applications of Lipatov's high energy effective \mbox{action} to NLO BFKL jet
phenomenology}

\author{{\slshape Martin Hentschinski$^1$, Beatrice Murdaca$^{2}$, Agust{\' i}n Sabio Vera$^1$}\\[1ex]
$^1$Instituto de F{\' i}sica Te{\' o}rica  
UAM/CSIC,   C/ Nicol\'as Cabrera 13-15,   \\ Universidad Aut{\' o}noma de Madrid, Cantoblanco, E-28049 Madrid, Spain
\\ \\
$^2$ Dipartimento di Fisica, Universit\`a della Calabria, 
and  \\Istituto Nazionale di Fisica Nucleare, Gruppo collegato di Cosenza, \\
I-87036 Arcavacata di Rende, Cosenza, Italy
}

\contribID{xy}

\doi  

\maketitle

\begin{abstract}
  We report on recent progress in the evaluation of next-to-leading order (NLO)
  observables using Lipatov's QCD high energy effective action. We
  calculate both real and virtual corrections to the quark induced
  forward jet vertex at NLO, making use of a new regularization method and a
  subtraction mechanism.  As a new result we determine the real part
  of the NLO Mueller-Tang impact factor which is the only missing
  element for a complete NLO BFKL description of dijet events with a
  rapidity gap.
\end{abstract}

\vspace{-12cm}
\begin{flushright}
  IFT-UAM/CSIC-12-54 \\
  LPN12-054 
\end{flushright}
\vspace{10.3cm}
\section{Introduction}

Due to its large center of mass energy the LHC provides an ideal
opportunity to test BFKL-driven observables \cite{Fadin:1975cb,
  Balitsky:1978ic}.  Among them we find  both central production processes,
such as heavy quark production (see {\it e.g.} \cite{Chachamis:2009ks}
) and forward production of different systems  such as high $p_T$ jets, heavy quark pairs \cite{Salas:prep} or Drell-Yan pairs \cite{Hautmann:2012sh, Hentschinski:prep}.
In the case of two tagged forward/backward jets there might be a rapidity gap between them (`Mueller-Tang') or not (`Mueller-Navelet'). 
These jet events allow then to test the forward (Mueller-Navelet) and non-forward (Mueller-Tang) BFKL kernel.
 While Mueller-Navelet jet events are currently one of
the few examples where a complete description at next-to-leading
logarithmic (NLL) accuracy exists~\cite{Fadin:1998py,
  Bartels:2001ge,Bartels:2002yj,Colferai:2010wu,
  Caporale:2011cc, Ivanov:2012ms}, for  Mueller-Tang jets we so far have at NLO accuracy  the non-forward BFKL kernel~\cite{Fadin:2005zj}, while  impact factors are known only to leading order (LO). The limitation to LO impact factors is currently one of the main drawbacks of BFKL phenomenology. 
 A promising tool to overcome this
limitation is given by Lipatov's effective action~\cite{Lipatov:1995pn}. So far this action has been mainly applied for
the determination of LO transition kernels~\cite{Hentschinski:2009ga,Hentschinski:2009cc,Hentschinski:2008rw,Hentschinski:2008im}. In this contribution  we show that it can be further used to calculate NLO
correction. In particular, we re-derive the NLO Mueller-Navelet
quark-jet impact factor and determine the missing real NLO correction
to the Mueller-Tang quark-initiated jet impact factors. For details we refer to~\cite{Hentschinski:2011tz, Hentschinski:prep2}.

\section{The high energy effective action}
\label{sec:LODYimpact}
 
   The
effective action adds to the QCD action an induced term, $
S_{\text{eff}} = S_{\text{QCD}} + S_{\text{ind.}}$, which describes
the coupling of the reggeized gluon field $A_\pm(x) = -i t^a A_\pm^a(x)
$ to the usual gluonic field $v_\mu(x) = -it^a v_\mu^a(x)$. This induced term
reads
\begin{align}
\label{eq:1efflagrangian}
  S_{\text{ind.}}[v_\mu, A_\pm]& = \int \! \text{d}^4 x \,
\tr\bigg[\bigg( W_+[v(x)] - A_+(x) \bigg)\partial^2_\perp A_-(x)\bigg]
\notag \\
&  \qquad  \qquad   \qquad
+\int \! \text{d}^4 x \, \tr\bigg[\bigg(W_-[v(x)] - A_-(x) \bigg)\partial^2_\perp A_+(x)\bigg]
.
\end{align}
The infinite number of couplings of the gluon field to the reggeized
gluon field are encoded in two functionals $W_\pm[v]  =
v_\pm \frac{1}{ D_\pm}\partial_\pm $ where $ D_\pm = \partial_\pm + g v_\pm$.
Note that  the reggeized gluon fields are special in the sense that they are
invariant under local gauge transformations, while they transform
globally in the adjoint representation of the SU$(N_c)$ gauge
group. In addition, strong ordering of longitudinal momenta in high
energy factorized amplitudes leads to the  kinematical
constraint of the reggeized gluon fields,
\begin{align}
  \label{eq:constraint}
\partial_+ A_-(x)  =  \partial_- A_+(x) = 0,
\end{align} 
which is always implied. 
Quantization of the gluonic field requires to add  gauge fixing and ghost terms, which we have included in  the QCD action.
\begin{figure}[htb]
  \centering
   \parbox{.7cm}{\includegraphics[height = 1.8cm]{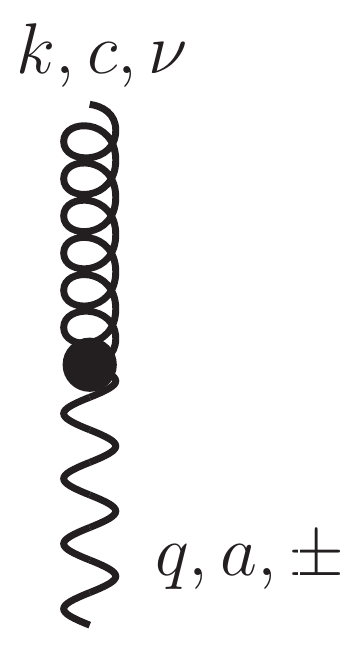}} $=  \displaystyle 
   \begin{array}[h]{ll}
    \\  \\ - i{\bm q}^2 \delta^{a c} (n^\pm)^\nu,  \\ \\  \qquad   k^\pm = 0.
   \end{array}  $ 
 \parbox{1.2cm}{ \includegraphics[height = 1.8cm]{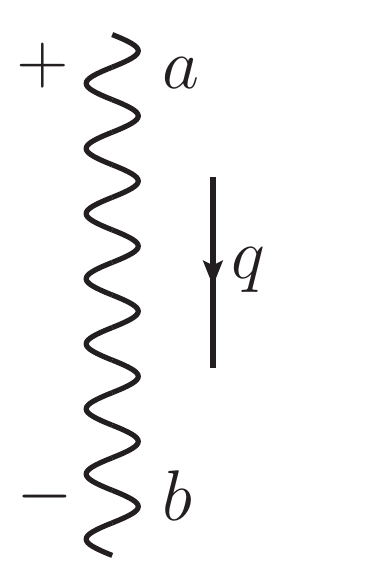}}  $=  \displaystyle    \begin{array}[h]{ll}
    \delta^{ab} \frac{ i/2}{{\bm q}^2} \end{array}$ 
 \parbox{1.7cm}{\includegraphics[height = 1.8cm]{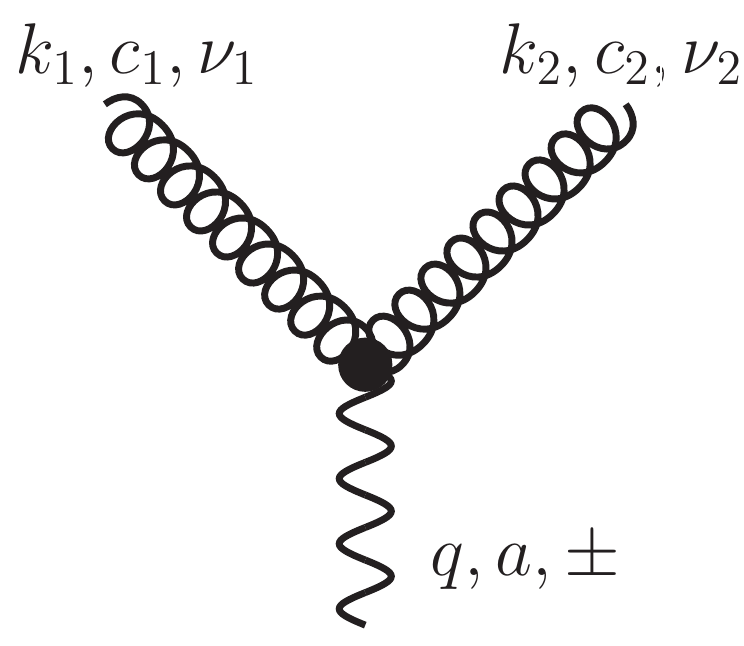}} $ \displaystyle  =  \begin{array}[h]{ll}  \\ \\ g f^{c_1 c_2 a} \frac{{\bm q}^2}{k_1^\pm}   (n^\pm)^{\nu_1} (n^\pm)^{\nu_2},  \\ \\ \quad  k_1^\pm  + k_2^\pm  = 0
 \end{array}$
 \\
\parbox{3cm}{\center (a)} \parbox{3cm}{\center (b)} \parbox{5cm}{\center (c)}
  \caption{\small The direct transition vertex (a), the reggeized gluon propagator (b) and the  order $g$ induced vertex (c) }
  \label{fig:feynrules0p2}
\end{figure}
Feynman rules  have been derived in~\cite{Antonov:2004hh}. We show them using curly lines for the
conventional QCD gluon field and wavy (photon-like) lines for the
reggeized gluon field.  There exist an infinite number of higher order
induced vertices. For the present analysis only  the order $g$
induced vertex in fig.~\ref{fig:feynrules0p2}.c is needed. In the
determination of loop corrections  we must fix a regularization
of the light-cone singularity present in
fig.~\ref{fig:feynrules0p2}.c. As suggested by one of us in~\cite{Hentschinski:2011xg} this pole should be treated as a Cauchy
principal value.

\section{NLO quark jet impact factors}
\label{sec:quarkjet}
\begin{figure}[h]
  \centering
 \parbox{.3 \textwidth}{ \center \includegraphics[width = 3cm]{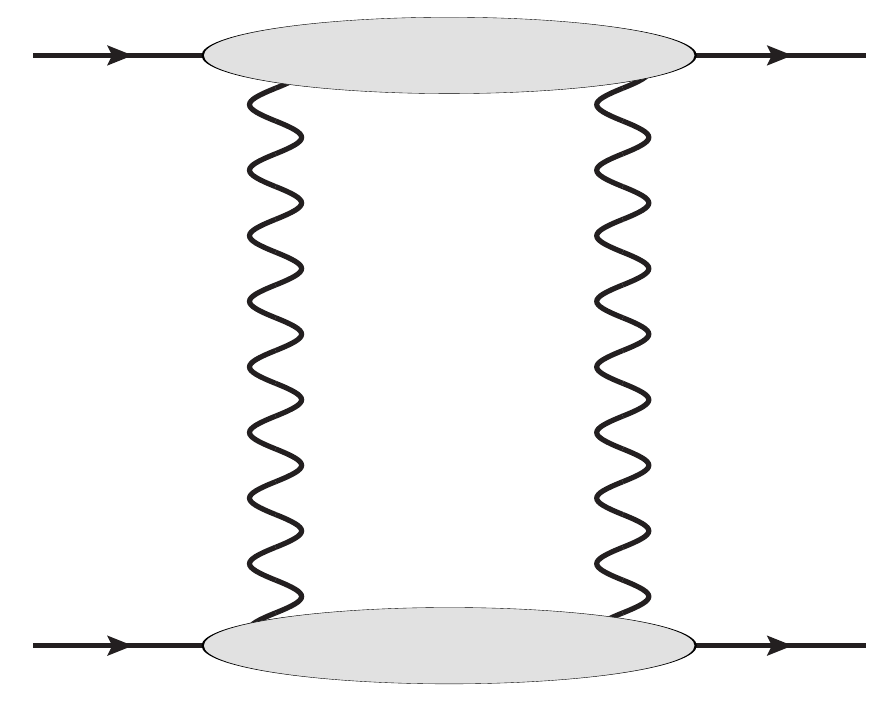}}   \parbox{.3 \textwidth}{\center \includegraphics[width = 3cm]{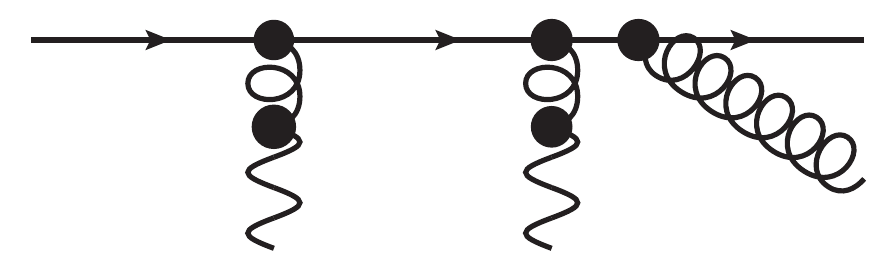}} 
 \parbox{.3 \textwidth}{\center \includegraphics[width = 3cm]{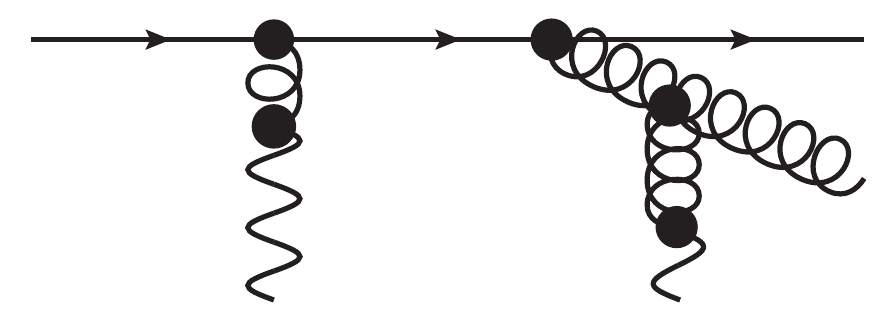}} 

  \parbox{.3 \textwidth}{\center (a)}   \parbox{.3 \textwidth}{\center (b)}  \parbox{.3 \textwidth}{\center (c)} 
  \caption{}
  \label{fig:mt}
\end{figure}

When calculating quantum corrections new divergences in longitudinal 
components appear. As it was demonstrated in~\cite{Hentschinski:2011tz,Chachamis:2012gh}  these can be regularized by deforming the light cone using a 
parameter $\rho$ which is considered in the limit $\rho \to \infty$. In this new setup, the 
Sudakov projections take place on the vectors $n_a = e^{-\rho} n^+ + n^-$ and $n_b = n^+ + e^{- \rho} n^-$. To 
obtain the virtual corrections we are seeking for it is needed to calculate the one-loop self energy corrections to the reggeized gluon
propagator. Diagrammatically, these are
\begin{align}
  \label{eq:1}
\parbox{.7cm}{\vspace{0.1cm} \includegraphics[height = 1.5cm]{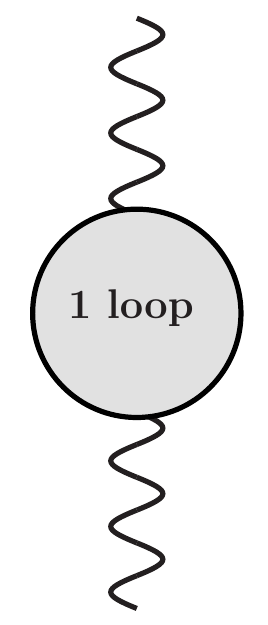}}
= 
    \parbox{.7cm}{\vspace{0.1cm} \includegraphics[height = 1.5cm]{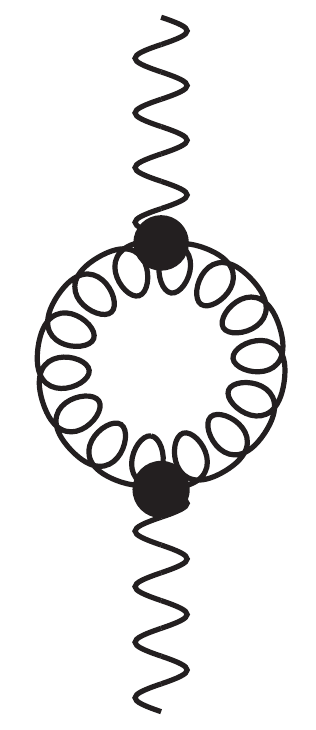}}
  + 
  \parbox{.7cm}{\vspace{0.1cm} \includegraphics[height = 1.5cm]{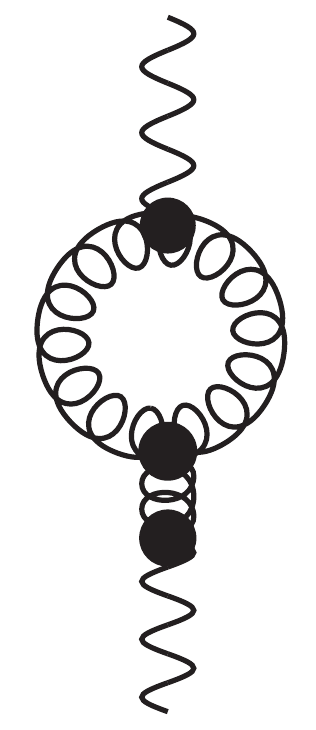}} 
 +
  \parbox{.7cm}{\vspace{0.1cm} \includegraphics[height = 1.5cm]{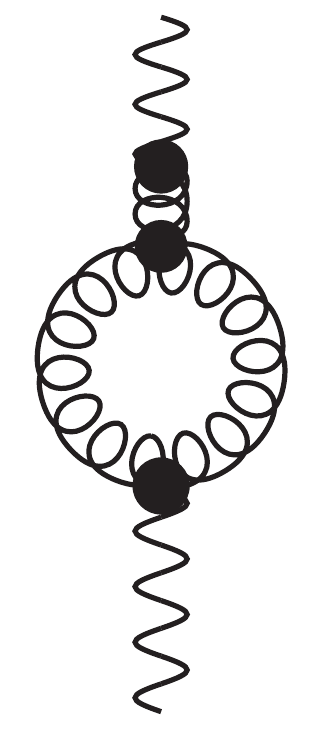}} 
 +
  \parbox{.7cm}{\vspace{0.1cm} \includegraphics[height = 1.5cm]{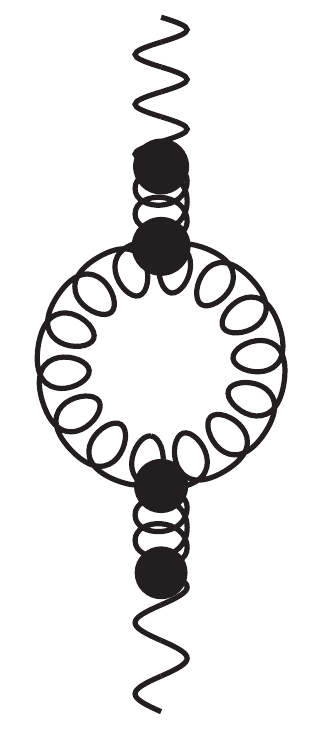}} 
 +
  \parbox{.7cm}{\vspace{0.1cm} \includegraphics[height = 1.5cm]{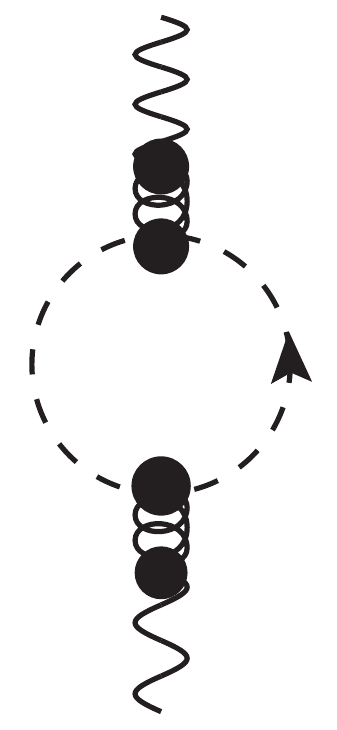}}
+
  \parbox{.7cm}{\vspace{0.1cm} \includegraphics[height = 1.5cm]{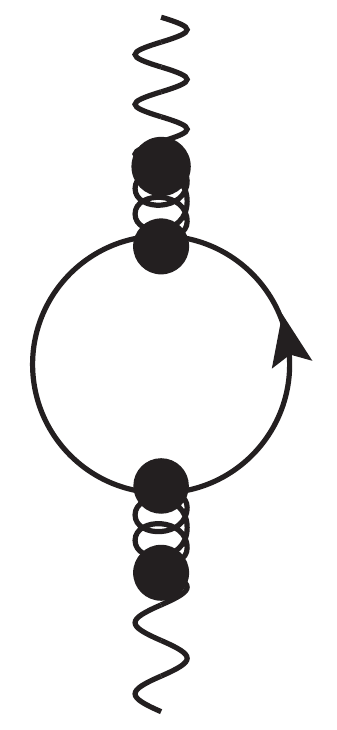}}.
\end{align}
The 1-loop corrections to the quark-quark-reggeized gluon vertex are
\begin{align}
  \parbox{1cm}{\includegraphics[width = 1cm]{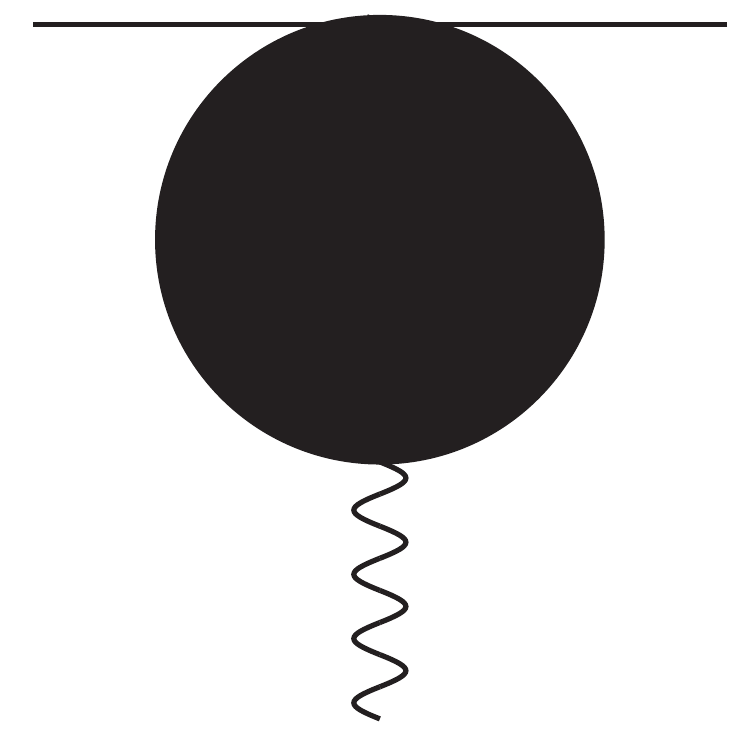}}  =
\parbox{1cm}{\includegraphics[width = 1cm]{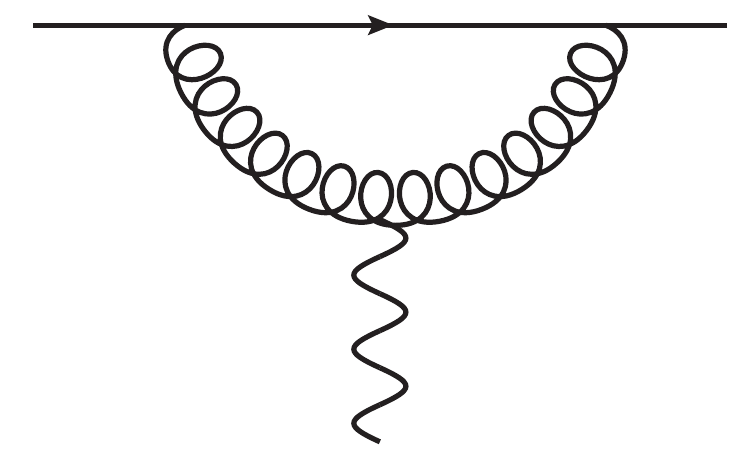}}
  + 
  \parbox{1cm}{\includegraphics[width = 1cm]{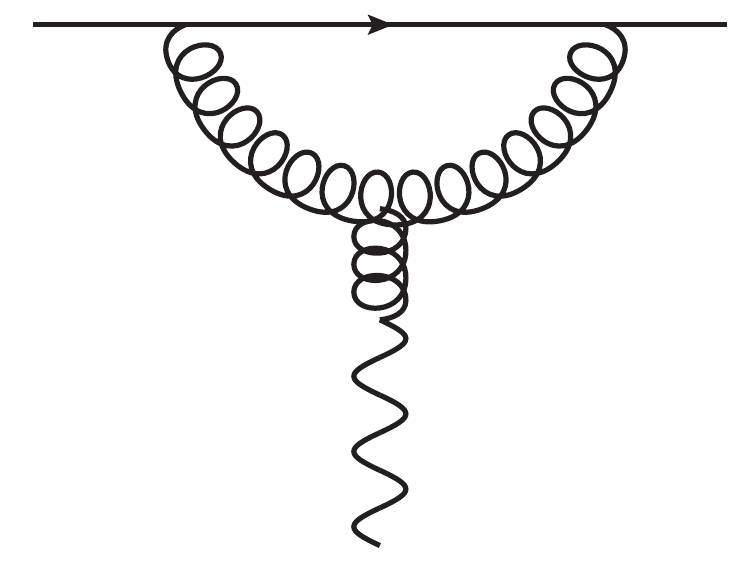}} 
  +
  \parbox{1cm}{\includegraphics[width = 1cm]{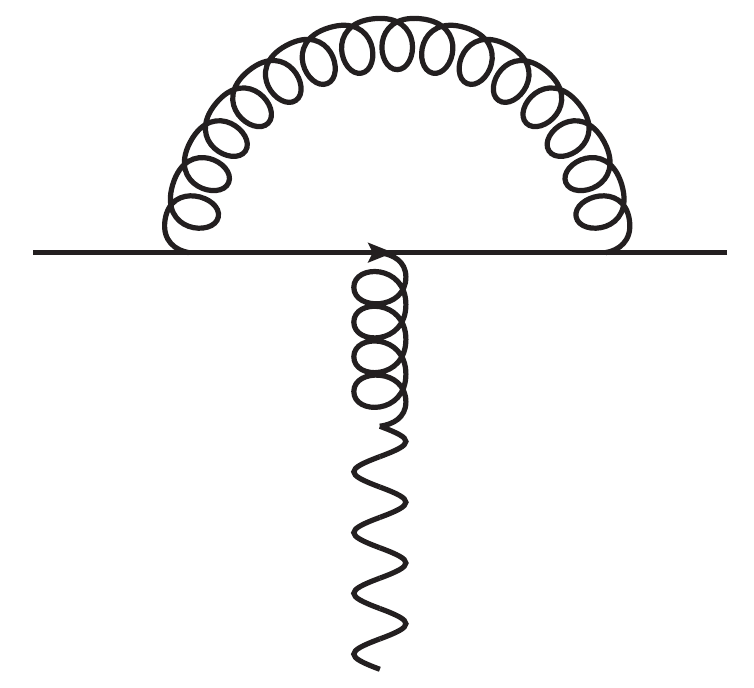}}
 + 
\parbox{1cm}{\includegraphics[width = 1cm]{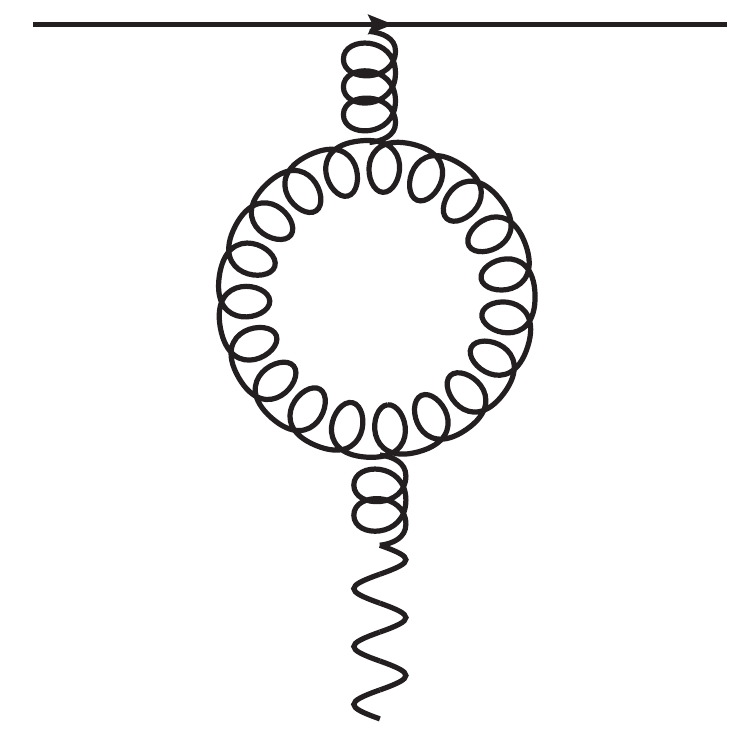}} 
+
\parbox{1cm}{\includegraphics[width = 1cm]{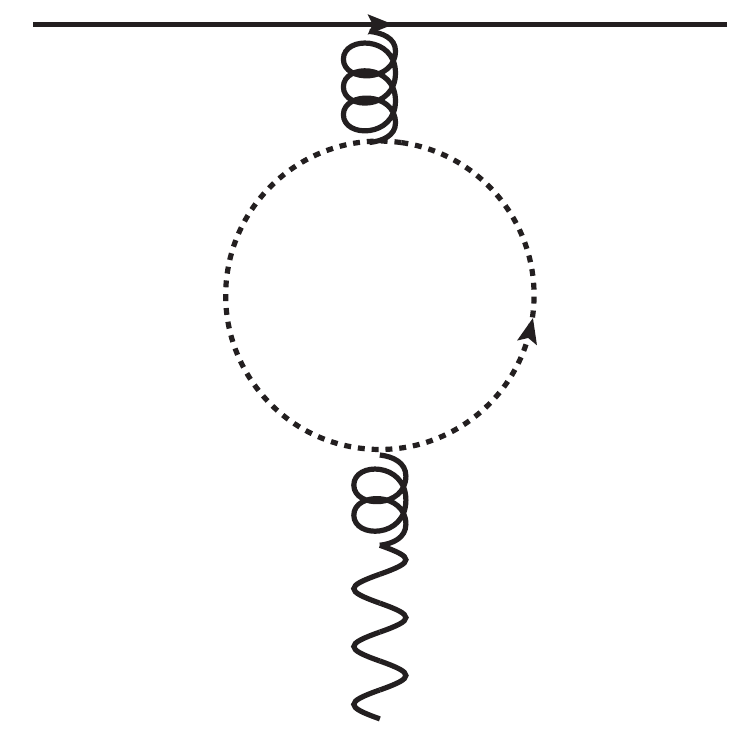}}
+
\parbox{1cm}{\includegraphics[width = 1cm]{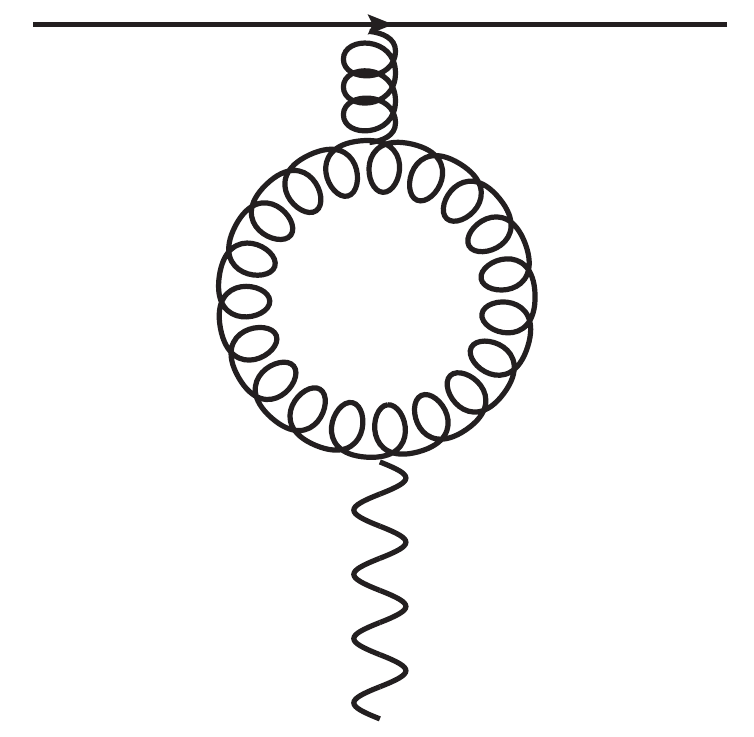}} 
+
\parbox{1cm}{\includegraphics[width = 1cm]{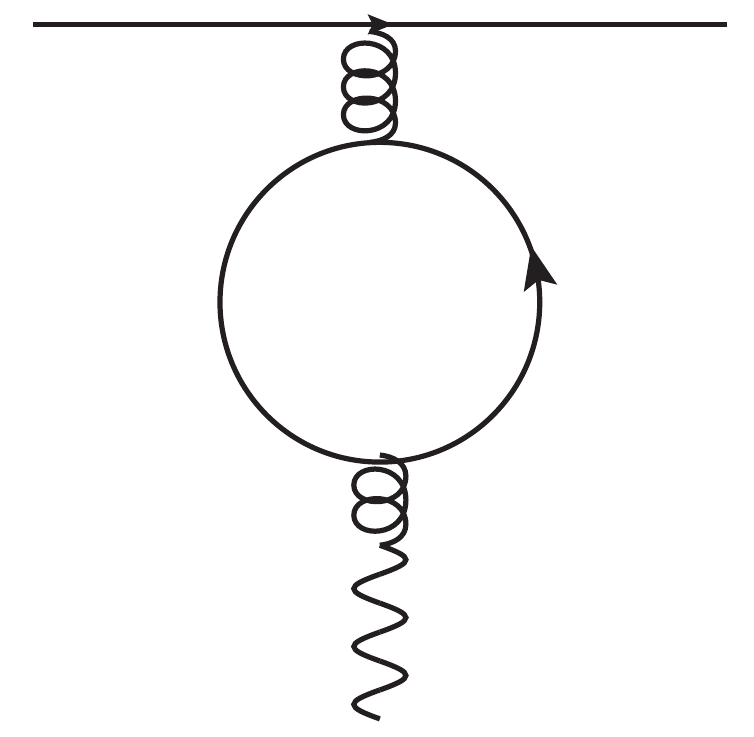}},
\end{align}
from which it is needed to subtract the factorizing contribution
\begin{align}
   \parbox{1cm}{\includegraphics[width = 1cm]{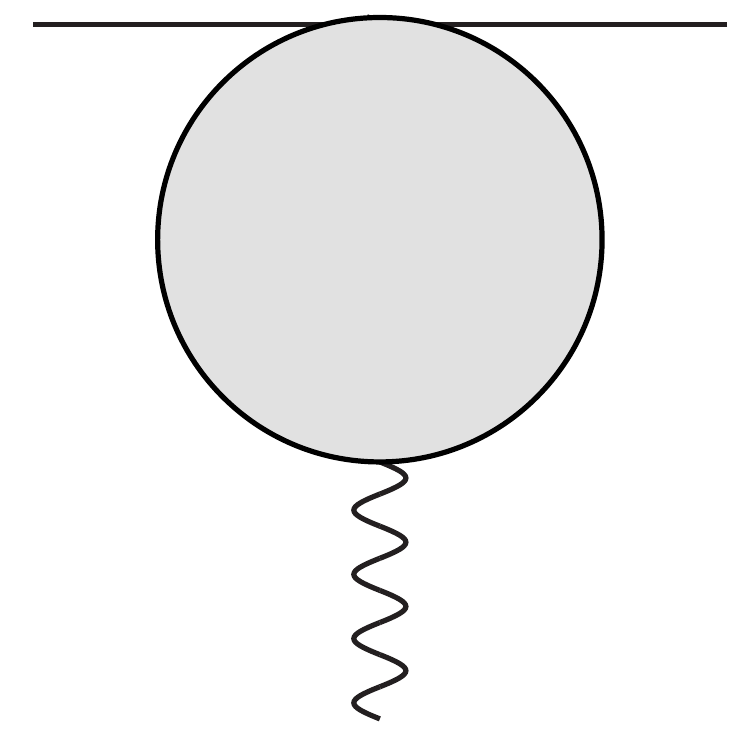}} &= 
\parbox{1cm}{\includegraphics[width = 1cm]{hentschinski_martinimpaamp.pdf}} 
- \parbox{1cm}{\includegraphics[width = 1cm]{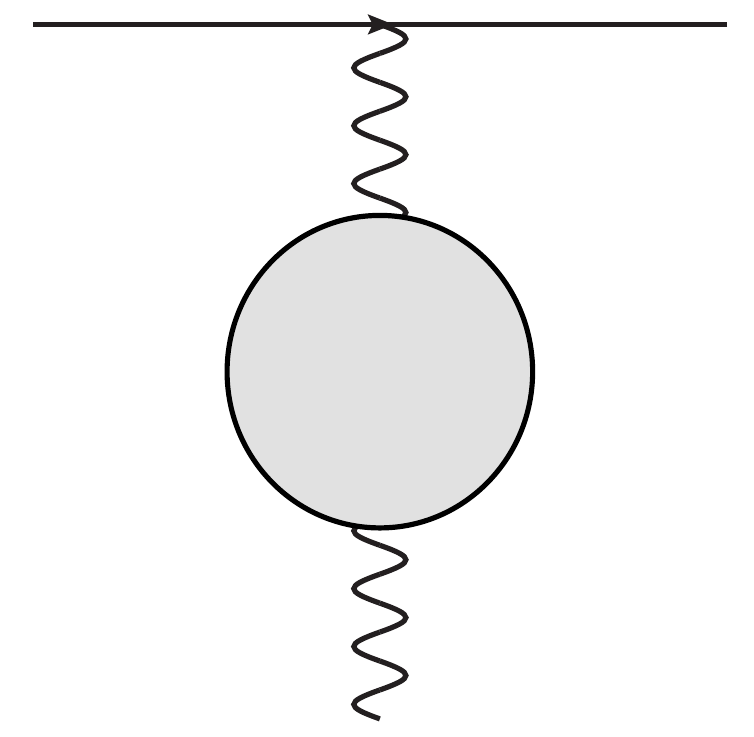}}.
\end{align}
The one-loop quark-quark scattering amplitude in the high energy limit is then given by the following sum
\begin{align}
\parbox{1.5cm}{\includegraphics[width = 1.5cm]{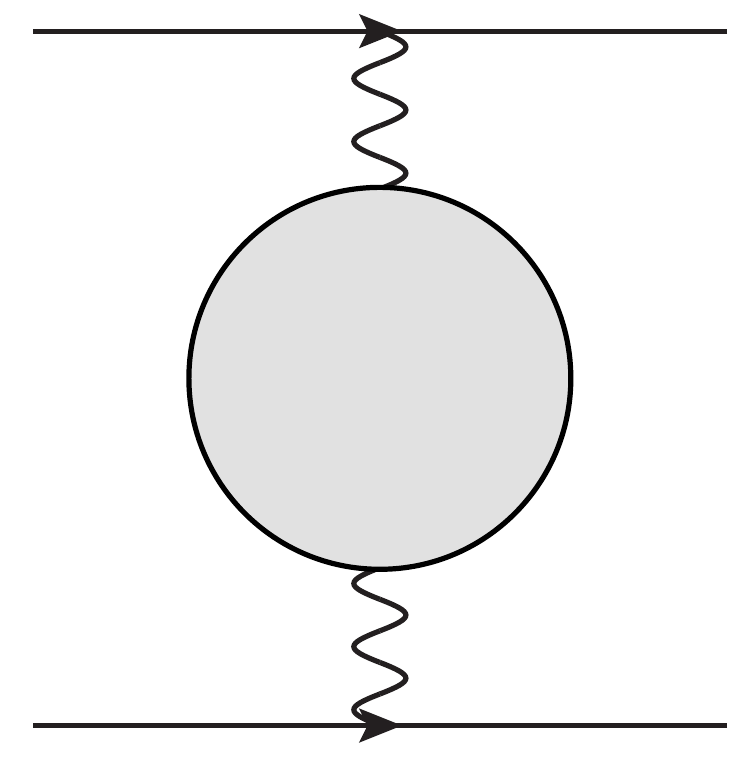}} 
+
  \parbox{1.5cm}{\includegraphics[width = 1.5cm]{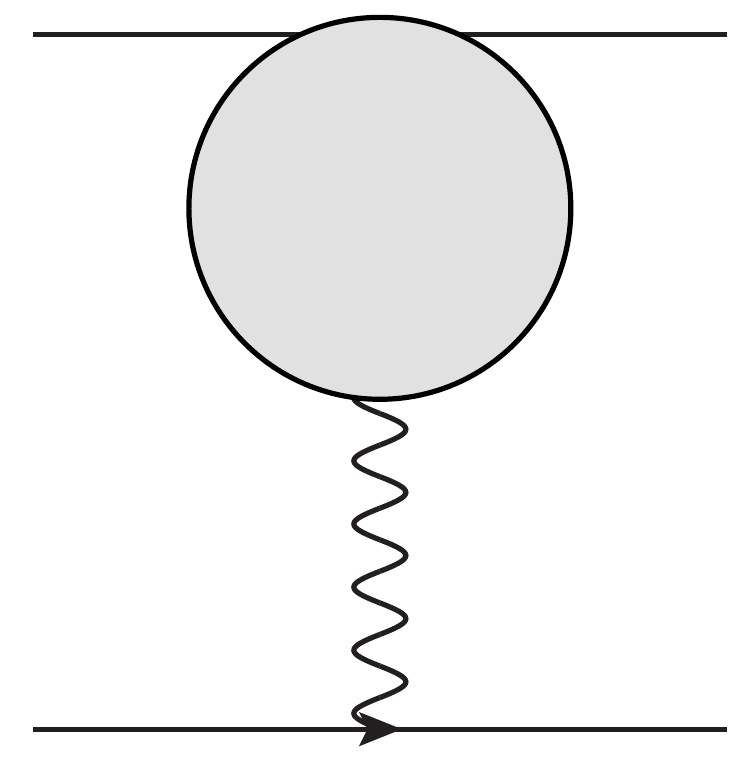}} 
+ 
 \parbox{1.5cm}{\includegraphics[width = 1.5cm]{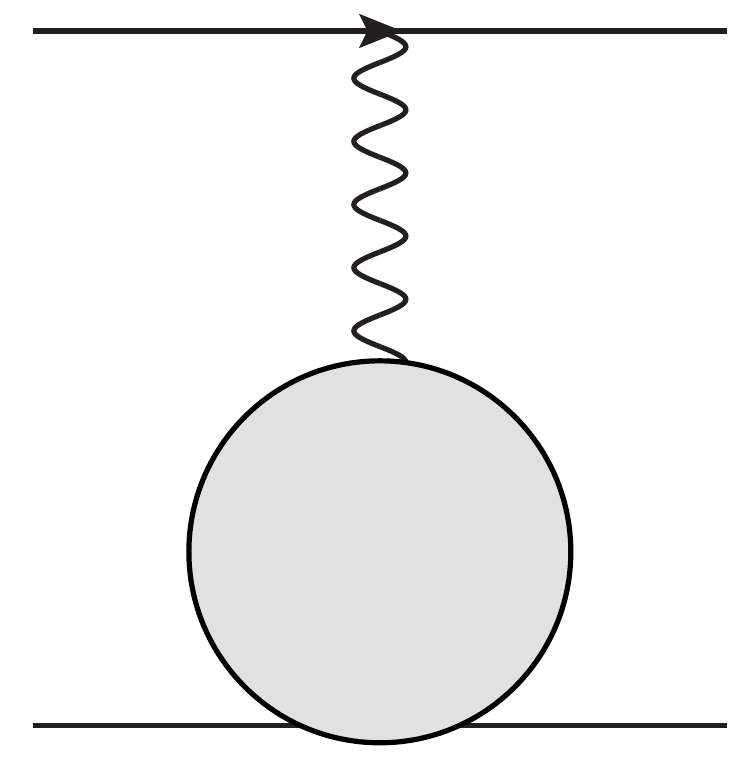}}. 
\end{align}
The sum of this three contributions is finite in the limit $\rho \to \infty$ and the dependence on the regulator vanishes.  The result is  in agreement with  calculations  using more standard techniques, performed in~\cite{Fadin:1993qb} and confirmed
in~\cite{DelDuca:1998kx}.  A similar result holds for the real
corrections, see~\cite{Hentschinski:2011tz} for details. All of these results are needed for Mueller-Navelet jets. The determination of the
Mueller-Tang impact factor requires to consider diagrams where two
reggeized gluons couple to the quark induced jet, see Fig.~\ref{fig:mt}.a. Due to the condition Eq.~(\ref{eq:constraint}),  the 
integration over the minus component of the  loop momentum of the reggeized
gluon loop is  absorbed into the definition of the impact factor. With  the  virtual NLO corrections already known~\cite{Fadin:1999df}, we focus on the real  NLO corrections.
The relevant  diagrams  split into two groups: the two reggeized gluon state couples either  to a single parton (Fig.~\ref{fig:mt}.b) or to two different  partons  (Fig.~\ref{fig:mt}.c). While the integration over the longitudinal loop momentum is divergent for individual diagrams, this divergence is found to cancel for their  sum.

Our results show the use of the high energy effective action in the determination of higher order corrections in multi-Regge and quasi-multi-Regge kinematics. In addition to the calculations here presented, these methods have been successfully applied to the  determination of the quark contributions to the two-loop gluon Regge trajectory~\cite{Chachamis:2012gh}. Determination of the gluonic NLO corrections to jet impact factor and gluon trajectory are currently in progress~\cite{Hentschinski:prep3}.

\section*{Acknowledgments}
We are grateful for financial support from the German Academic
Exchange Service (DAAD), the MICINN under grant FPA2010-17747, the
Research Executive Agency (REA) of the European Union under the Grant
Agreement number PITN-GA-2010-264564 (LHCPhenoNet) and Comunidad de Madrid (HEPHACOS ESP-1473).
 

{\raggedright
\begin{footnotesize}



 \bibliographystyle{DISproc}
 \bibliography{hentschinski_martin_biblio2.bib}
\end{footnotesize}
}


\end{document}